\documentclass[prl,showpacs,showkeys,twocolumn,floatfix]{revtex4} 
\usepackage[]{epsfig,dcolumn}
\begin{document}
\title{A search for $J^{PC}=1^{-+}$ exotic mesons in the $\pi^- \pi^- \pi^+$ and $\pi^- \pi^0 \pi^0$ systems}
\author{A.~R.~Dzierba}
\author{R.~Mitchell}
\author{A.~P.~Szczepaniak}\altaffiliation[Also with the ]{Nuclear Theory Center, Indiana University}
\author{M.~Swat}
\author{S.~Teige}
\affiliation{Department of Physics, Indiana University, Bloomington, IN 47405.}
\date{\today}

\begin{abstract}
A partial wave analysis (PWA) of the  $\pi^- \pi^- \pi^+$ and $\pi^- \pi^0 \pi^0$ systems 
produced in the reaction $\pi^- p \to (3\pi)^-p$ at 18~GeV/$c$ was carried
out using an \emph{isobar} model assumption.  This analysis is based on 3.0M
$\pi^- \pi^0 \pi^0$ events and 2.6M  $\pi^- \pi^- \pi^+$  events and
shows production
of the $a_2(1320)$, $\pi_2(1670)$ and $\pi(1800)$ mesons.  
An earlier analysis of 
250K $\pi^- \pi^- \pi^+$ events from the same experiment
 showed possible evidence for a $J^{PC}=1^{-+}$ exotic meson
with a mass of 1.6~GeV/$c^2$ decaying into $\rho \pi$.
In this analysis of a higher statistics sample of the $(3\pi)^-$ system in two
charged modes
we find no evidence of an exotic meson.

\end{abstract}

\maketitle

Quantum chromodynamics (QCD) predicts a spectrum of \emph{hybrid} mesons beyond the
$q \bar q$ bound states of the conventional quark model in which the gluons
binding the quarks manifest their degrees of freedom.  Lattice QCD and QCD-inspired
models predict that the gluonic field within the meson forms a flux tube for quark
separation greater than 1~fermi \cite{colin}.  When
the flux tube is in its ground state conventional mesons emerge.  When the flux-tube
is in its excited states hybrid mesons emerge.  For conventional mesons the 
spin ($J$), parity ($P$) and charge conjugation ($C$) 
 quantum numbers of
the $q \bar q$ system are those of a fermion--antifermion system: $\vec{J}=\vec{L}+\vec{S}$
where $\vec{L}$ is the angular momentum between the quarks and $\vec{S}$ is
the total quark spin; $P=(-1)^{L+1}$; and $C=(-1)^{L+S}$.  Thus $J^{PC}=0^{+-},
1^{-+}, 2^{+-}, \dots$ are not allowed.  These \emph{exotic} quantum numbers are
allowed when the quantum numbers of the excited flux-tube are included and are thus
a signature for hybrid mesons.  The fundamental interest in the spectrum and
other properties (decay modes and widths)
of hybrid mesons is that flux tubes are thought to be
responsible for the confinement of quarks and gluons in QCD. 

In 1998, the E852 collaboration reported evidence for the $\pi_1(1600)$, a 
 $J^{PC}=1^{-+}$ exotic hybrid meson with a mass of 1.6~GeV/$c^2$ decaying
 into $\rho \pi$.  The original publication \cite{Adams98} of this result was followed by another
 \cite{Chung02} giving more details of the analysis  based on 250,000 events
 of the reaction $\pi^- p \to \pi^- \pi^- \pi^+ p$ at 18~GeV/$c$ collected in the E852
 experiment at Brookhaven Lab in 1994. A search for this state by the VES collaboration
 in the reaction $\pi^-A \to \pi^- \pi^- \pi^+ A$ interactions at 37~GeV/$c$ \cite{VES} did not
 confirm this state.
   The partial wave analysis (PWA) procedure
 used to extract these results was based on the \emph{isobar} model assumption
 whereby the final state $3 \pi$ system is reached via an intermediate
 state of a di-pion resonance (\emph{e.g.}  
$ \sigma$, $f_0(980)$, $\rho(770)$,
 $f_2(1275)$, $\rho_3(1690)$) and a bachelor $\pi$.    
 
 \begin{figure}  
\centerline{\epsfig{file=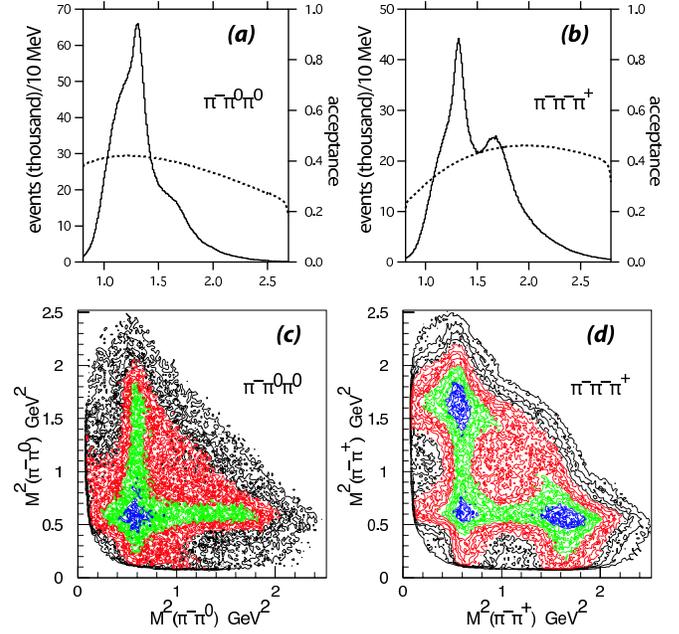,width=\columnwidth}}
\caption{
 Acceptance-uncorrected (a) $\pi^-\pi^0\pi^0$ and (b)  $\pi^-\pi^-\pi^+$
 effective mass distributions along with acceptance functions and
 the (c) $\pi^-\pi^0\pi^0$ and (d)  $\pi^-\pi^-\pi^+$ Dalitz plots
  for events in the $\pi_2(1670)$ mass region.
}\label{fig1}
\end{figure}

   This paper reports on the analysis of additional data collected in
 1995 including 3.0M events of the 
 reaction   $\pi^- p \to \pi^- \pi^0\pi^0 p$ and 2.6M events of the reaction
  $\pi^- p \to \pi^- \pi^- \pi^+ p$.  
   We compare the isobar model PWA of these two
    large data sets with each other and with the analysis presented
    in \cite{Adams98,Chung02}.  
    The comparison of the two
    $(3\pi)^-$ modes provides powerful cross checks.
Any resonance decaying to $(\rho\pi)^-$ should decay equally to 
$\rho^-\pi^0 \to (\pi^-\pi^0)\pi^0$ and $\rho^0\pi^- \to (\pi^+\pi^-)\pi^-$
 and thus appear with equal probabilities in the two
modes.  Similarly, any resonance decay to $f_2\pi^-$ ought to appear twice as
often in $(\pi^+\pi^-)\pi^-$ as $(\pi^0\pi^0)\pi^-$.  And these same isospin rules dictate
relationships between the two modes for all other decay sequences.  Since
the $\pi^- \pi^- \pi^+$ and $\pi^- \pi^0 \pi^0$ modes rely on different elements of the
detector, 
any misunderstandings in the acceptance would affect the two
modes differently and inconsistencies would result.  As the results will
show, this is not the case.

The data reported on in this paper were collected in the E852 experiment at
 the Alternating Gradient Synchrotron (AGS) at
 Brookhaven Laboratory (BNL) using the multi-particle spectrometer (MPS)
 augmented with a forward electromagnetic calorimeter (LGD) consisting of
 3000~lead glass blocks.
 An 18.3~GeV/$c$ $\pi^-$ beam was incident on a 
  liquid hydrogen target.  Details about the E852 experiment are given 
  elsewhere \cite{equip,LGD}.  

 \begin{figure}  
\centerline{\epsfig{file= 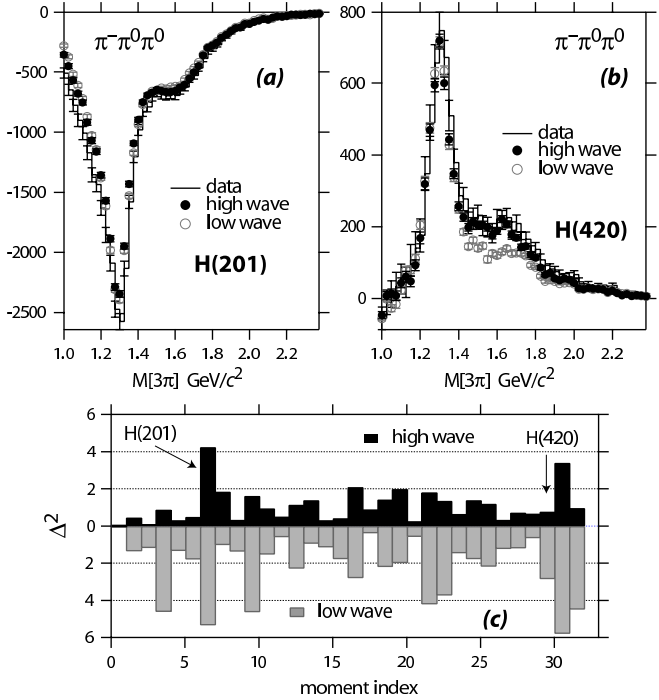,width=\columnwidth}}
\caption{
Comparison of the (a) H(201) and (b) H(420)
 moments as computed directly from data and 
from PWA fits for the low and high wave sets for the $\pi^-\pi^0\pi^0$
channel.  In (c) the $\Delta^2$ (differences squared
divided by errors squared summed over all mass bins
and divided by the number of mass bins) for various
moments. 
}\label{fig2}
\end{figure}

\begin{figure}  
\centerline{\epsfig{file= 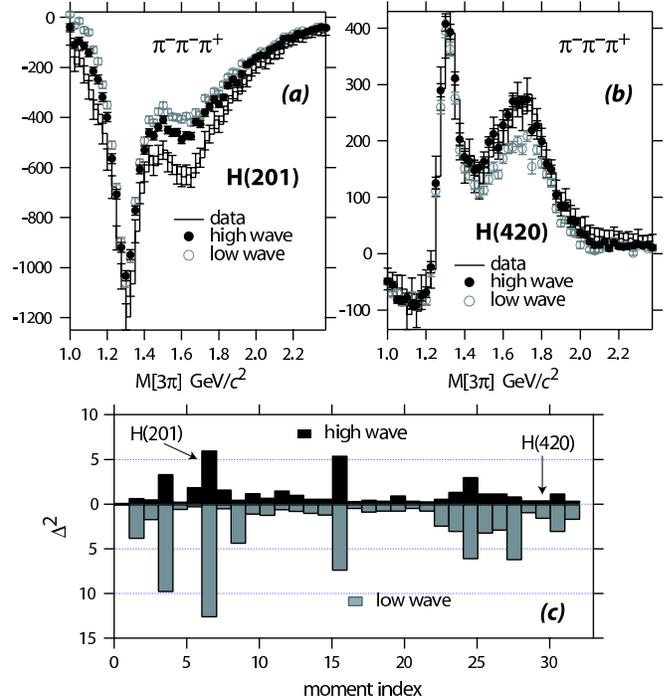,width=\columnwidth}}
\caption{
Comparison of the (a) H(201) and (b) H(420)
 moments as computed directly from data and 
from PWA fits for the low and high wave sets for the $\pi^-\pi^-\pi^+$
channel.  In (c) the $\Delta^2$ (differences squared
divided by errors squared summed over all mass bins
and divided by the number of mass bins) for various
moments.  
}\label{fig3}
\end{figure}

The 1995 run of E852 collected approximately 146M (265M) triggers requiring
one (three) forward going charged particles. For those triggers requiring only
one forward track, it was also required that the trigger processor 
\cite{LGD}
detected an energy deposition pattern in the LGD consistent with an effective mass
greater than a single $ \pi^0$. 
The kinematic fitting program SQUAW \cite{SQUAW} was used in identifying 
approximately 73M (79M) reconstructed events that were
consistent with the topology $ \pi^- p \rightarrow (3\pi)^- p$.
This constrained fit used the measured track and beam momenta, the measurements
from the LGD and constraints on the recoil particle mass ($ = m_p$) and di-photon
effective masses (two pairs $=m_{\pi}$) to refine the measured final state particle
4-momenta. In the $ \pi^- \pi^0 \pi^0 $ topology, 13.8M events survived this selection while
16.8M events of the $ \pi^- \pi^- \pi^+ $ topology survived.  Further cuts on fit confidence
level, vertex position and correlation between the $3\pi$ system and a recoil track
produced the final data set analyzed here:  3,025,980 $ \pi^- \pi^0 \pi^0 $events and
2,585,852  $ \pi^- \pi^- \pi^+ $events.

  \begin{figure}  
\centerline{\epsfig{file= 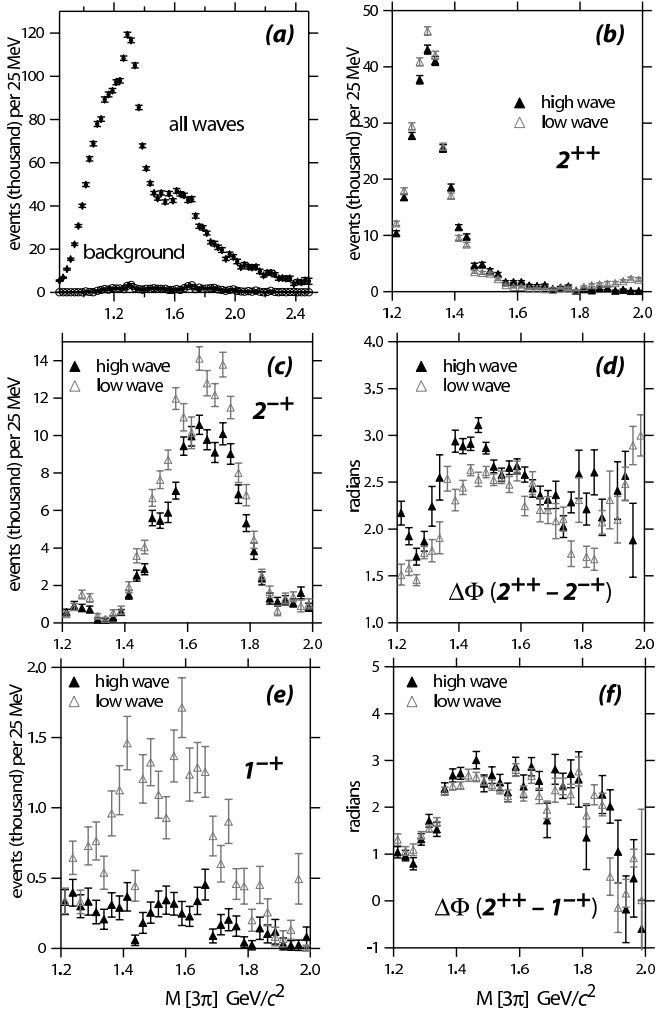,width=\columnwidth}}
\caption{
PWA results for the $\pi^-\pi^0\pi^0$ channel as a
function of $3\pi$ effective mass:  (a) sum of all waves and the background
wave; (b) $2^{++}$ wave; (c) $2^{-+}$ wave; (d) $\Delta \Phi(2^{++}-2^{-+})$; 
(e) $1^{-+}$ wave; (f) $\Delta \Phi(2^{++}-1^{-+})$.  For (b) through (f) the results
for the low wave and high wave sets are shown.
}\label{fig4}
\end{figure}

\begin{figure}  
\centerline{\epsfig{file= 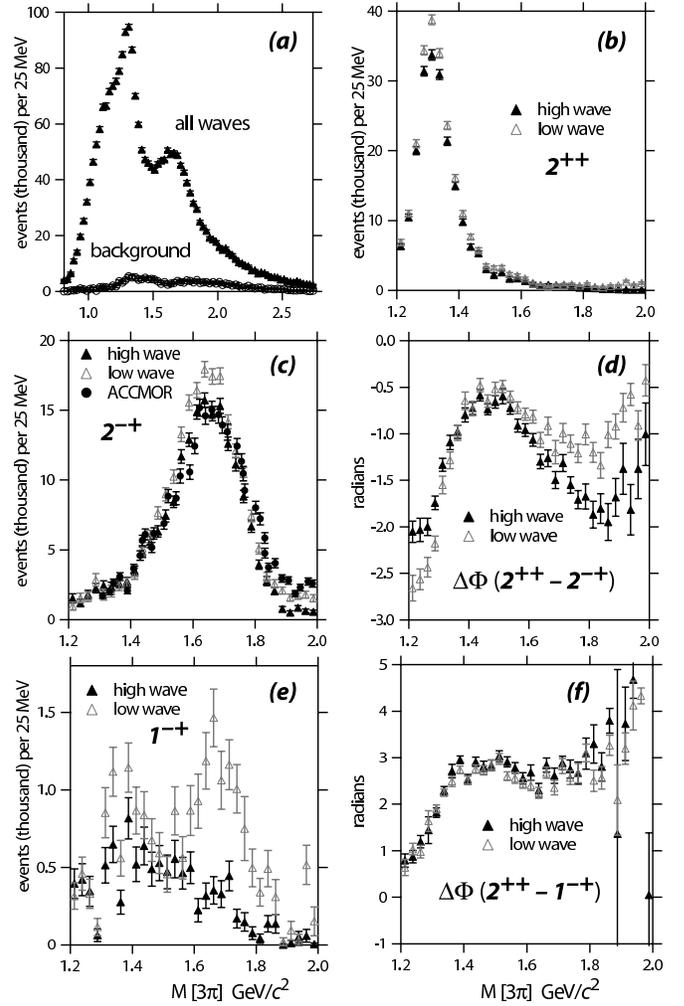,width=\columnwidth}}
\caption{
PWA results for the $\pi^-\pi^-\pi^+$ channel as a
function of $3\pi$ effective mass:  (a) sum of all waves and the background
wave; (b) $2^{++}$ wave; (c) $2^{-+}$ wave; (d) $\Delta \Phi(2^{++}-2^{-+})$; 
(e) $1^{-+}$ wave; (f) $\Delta \Phi(2^{++}-1^{-+})$.  For (b) through (f) the results
for the low wave and high wave sets are shown.
 }\label{fig5}
\end{figure}

 Figures~\ref{fig1} (a) and (b) show the acceptance uncorrected
 $\pi^- \pi^0 \pi^0$ and $\pi^- \pi^-\pi^+ $
effective mass distributions respectively.  
Enhancements in the $a_1(1260)$, $a_2(1320)$
and $\pi_2(1670)$ mass regions are observed.
The acceptance in $3\pi$ effective mass is also shown in 
Figures~\ref{fig1} (a) and (b).
Dalitz plots for the $\pi^- \pi^0 \pi^0$ and $\pi^- \pi^-\pi^+ $
 systems in the $\pi_2(1670)$ mass region are shown in
 Figures~\ref{fig1} (c) and (d) respectively.  These
Dalitz plots clearly show rich structure and bands
corresponding to decays into $\pi \rho(770)$
and $\pi f_2(1275)$, motivating application of the isobar model.

The PWA software used in this analysis consisted of programs developed
at BNL and  Indiana University (IU)
\cite{IUPWA}.  The formalism
is similar to that used in \cite{Adams98,Chung02}
but the details of handling the data and doing the the PWA
fits differ.  The IU software was optimized for running on a 200-processor
computer cluster (AVIDD) \cite{AVIDD} allowing systematic studies involving
 many PWA fits, varying isobar parameters
and using different wave sets. Results from the two programs
were compared with each other and found to be consistent.
The IU programs were also used to 
analyze the data presented in \cite{Adams98,Chung02}
and produced consistent results.


The analysis is performed in the Gottfried-Jackson frame. Within the isobar formalism each partial wave is characterized by $J^{PC}M^{\epsilon},(SL)$, where $J$, $P$ and $C$ are the spin, parity and charge conjugation of the $3\pi$ system, respectively, $M$ is the spin projection along the $z$ axis  and $\epsilon$ represents symmetry of the $3\pi$ system under reflection in the production plane. Finally $S$ is the spin of the isobar and $L$ is the relative orbital angular momentum between the isobar and the bachelor $\pi$. 
Except for the $S=0$ wave, all $\pi\pi$ resonances are parametrized using relativistic Breit-Wigner functions with Blatt-Weisskopf factors.

Following \cite{Daum81} a sufficient set of partial waves was determined
by sequentially removing waves from a parent set containing waves
with $ J\leq 4$, $M \leq 1$ and $S \leq 3$ and examining the resulting change
in likelihood. If no significant change in likelihood was obtained, the wave
was removed.
The exotic $1^{-+}$ partial waves were kept even though they would have been
chosen for removal by the above criteria -- the existence of signals in these
waves is discussed below.  Based on these criteria the compiled set 
includes 35 waves  and a background wave-- we refer to this as the \emph{high wave} set.
In comparison, the wave set used in the analysis described in \cite{Chung02}
included 20 waves and a background wave -- we refer to this as the \emph{low wave} set.
The 35 wave set included three additional
$ 1^{++}$, seven additional $ 2^{-+}$ and seven additional $ J \geq 3 $ waves
This high wave set did not include two negative reflectivity partial
waves ($1^{++}1^-,(10)$ and $2^{-+}1^-,(20)$) that were included
in the low wave set PWA. 

Another arbiter of wave set sufficiency is the comparison of moments, $H(LMN)$, 
of the $D^L_{MN}(\Omega)$ functions
as calculated directly from the data and as calculated using the
results of PWA fits.  We define $H(LMN) =
 \int  I(\Omega)D^L_{MN}(\Omega)d\Omega$ where $\Omega$
 represents the Euler angles of the $3\pi$ system \cite{Ascoli}
 and the intensity $I(\Omega)$ is determined directly from experiment
 or computed using the results of the PWA fits.
In Figure~\ref{fig2}
we show the H(201) and H(420) moment comparisons
as a function of $3\pi$ mass for the
$\pi^-\pi^0\pi^0$ channel for low wave and high wave set PWA fits.
We also show the difference between data and PWA calculations
of the moments as  $\Delta^2$ (summing differences squared
divided by errors squared summed over all mass bins
and divided by the number of mass bins) for various
moments.  
Similar plots are shown in Figure~\ref{fig3} for the $\pi^- \pi^-\pi^+ $ channel.
The moments calculated using the PWA results from the high wave set
have better agreement with experimental moments.  But there are
moments (such as H(201) - shown in part (a) of Figure~\ref{fig3})
for which agreement is not achieved and this is most likely due to the inherent
inadequacy of the isobar model in describing the underlying production
mechanism.

Figures~\ref{fig4} and \ref{fig5} show the PWA fit results for the $\pi^-\pi^0\pi^0$
and $\pi^-\pi^-\pi^+$ channels respectively as a function of the $3\pi$ effective
mass. 
 In part(a) of each figure the sum of the waves 
is shown along with the background wave.  The background wave is added
incoherently with the other waves and is isotropic in all decay angles.
The fitting procedure constrains the sum of all waves to describe the
observed $3\pi$ effective mass.  
Parts (b), (c) and (d) of each figure show the
mass dependence of the $2^{++}1^+,(12)$ and $2^{-+}0^+,(20)$ waves and their phase
difference.  Results are shown for both the low wave and high wave set.  
To within 10\% the
acceptance corrected yields of $a_2^-$ in the
two $3\pi$ modes are equal, consistent with expected decays into $\rho^- \pi^0$
and $\rho^0 \pi^-$.  And the yield for $\pi_2^-$ is higher by a factor of two
for the $f_2(1275) \pi^-$ modes when $f_2 \to \pi^+ \pi^-$ compared to $f_2 \to \pi^0 \pi^0$,
consistent with expectations from isospin. The $2^{++}$ and $2^{-+}$ waves and their phase
difference are consistent with two interfering Breit-Wigner line shapes.
For the $2^{-+}$ wave in the  $\pi^-\pi^-\pi^+$ channel (Figure~\ref{fig5}(c)) we 
also show the published  $2^{-+}$ wave from the PWA of diffractive $3\pi$
production in $\pi^-p$ interactions at 63 and 94~GeV/$c$ from the
ACCMOR collaboration \cite{Daum81}.

The $1^{-+}1^+,(11)$ wave and its interference
with the $2^{++}$ wave are shown in parts (e) and (f) of each figure.  When
the low wave set is used in the fit, an enhancement is observed
for the exotic $1^{-+}$ wave in the $3\pi$
mass region around 1.6~GeV/$c^2$, consistent with the observation of 
\cite{Adams98,Chung02} -- \emph{i.e.} we obtain the same results 
when we use the same wave set.
But the enhancement disappears when the
high wave set is used.  This
effect is observed for both $3\pi$ channels. 
 And the phase (see part (d)) of the
exotic wave relative to the dominant $2^{++}$ wave is similar for
both wave sets and dominated by the phase of the $a_2(1320)$.
The amplitudes of the negative reflectivity exotic waves show the same
behavior, disappearing as the fit improves.

 The masses and
widths of the $ \rho $ and the $f_2(1275)$ were varied and the  maximum likelihood
was achieved using current best values for these quantities \cite{PDG}.
Different parameterizations of the $ \pi\pi $ S-wave were also examined
\cite{AMP,oset,Chung02}.
While small changes in the likelihood were observed, no systematic changes in the
PWA results occurred.

The PWA results presented here were for data with $0.18 < |t| < 0.23$~(GeV/$c$)$^2$
where $t$ is the square of the momentum transferred  from the incoming
$\pi^-$ to the outgoing $(3\pi)^-$ system.  PWA fits were also carried out
for 12 other regions in $t$ and the results are similar to those presented
here.  

In summary, there is no
evidence  for an exotic $J^{PC}=1^{-+}$ meson in the $3\pi$ system
and in particular for the $\pi_1(1600)$.

\section {Acknowledgments}
This work was supported by grants from the United States Department of
Energy
(DE-FG-91ER40661/Task~D and DE-FG0287ER40365) and
 the National Science
Foundation (EIA-0116050) for AVIDD.


\begin{thebibliography}{99}
\bibitem{IUPWA} Details about the IU software and PWA can be found at
the website: dustbunny.physics.indiana.edu/3p\_paper.
\end{thebibliography}

\end{document}